\documentstyle[12pt]{article}
\documentstyle{fleqn}

\begin{document}
%\vfill\eject
\begin{flushright}
OU-HET 197
\linebreak
September 27, 1994
\end{flushright}
\vskip1.5cm
\begin{center}
{ \large
{\bf
Radiative Corrections to the Higgs Boson Decay
into
\linebreak
a Longitudinal $W$-Boson Pair
in a Two-Doublet Model
}}
\end{center}
%{\fontA \fox \par}
\vskip1.5cm
\begin{center}
{\large Shinya Kanemura and Takahiro Kubota}
\end{center}
\vskip1cm
\begin{center}
{\it Department of Physics,   Osaka University,
\linebreak
Toyonaka, Osaka 560, Japan}
\end{center}
\vskip1cm
\begin{center}
{\bf Abstract}
\end{center}
The decay rate of a (neutral) Higgs boson ($H$)
 going into a longitudinal $W$-boson pair is
calculated by including one loop radiative
corrections in a two-Higgs
doublet       model.  It is assumed that the
 Higgs boson $H$    is much heavier than the $W$
boson and a full  use has been made of the
equivalence theorem.
A possibility is explored extensively that the
precise measurement of the decay rate of $H$
could be useful as a  probe of  the other neutral
as well as charged Higgs bosons through
potentially important  radiative corrections.
It is pointed out that, for some choice of the
parameters, the radiative corrections are sizable
and that there exist a possibility of extracting
useful information on the other Higgs boson masses.
\pagebreak
\begin{center}
\large
{\bf 1. Introduction}
\end{center}

In the last two decades there have been a lot of
efforts  towards discovery of Higgs bosons both
theoretically and experimentally. (See Ref. [1]
for a review.) There is potentially a hope on the
theoretical side that,  even
before  direct detections of
 the Higgs  bosons, one might
be able to  get indirect signatures
 of  the existence of the Higgs particles: {\it i.e.,}
the study of radiative corrections and precision tests
 of electro-weak processes have been often motivated
to see low-energy manifestations of unknown very heavy
particles.

Veltman [2] has once argued that in the minimal
standard model with  a single  Higgs boson
doublet, the  Higgs boson mass
dependence of  radiative corrections would  be
at most  logarithmic on the one-loop level
and therefore the internal loop effects of
the Higgs boson
are rather  elusive.  Einhorn and Wudka [3] studied
radiative corrections to gauge boson propagators in a
general way and strengthened the Veltman's claim.
They realized that, due to the custodial $SU(2)$
symmetry in a single Higgs scheme [4], the Higgs mass
dependence  is reduced to a logarithmic one to all
orders in perturbation thoery. This fact is sometimes
referred to as Veltman's screening theorem.

The analysis of Einhorn and Wudka tells us
 that radiative corrections other than those to
vector-boson propagators
 may have power-type terms w.r.t. Higgs masses.
 It is also clear  that, in a
model where the custodial symmetry is not respected,
  there is no compelling reason a priori
that Higgs bosons
 do not produce power-like corrections to gauge boson
 propagators.  Two-Higgs doublet models belong to
 such examples and  the calculation performed  by
Toussaint [5] shows in fact that  the screening theorem
does not apply to the two-doublet model.
This fact prompted  several authors [6-7] to study
radiative corrections to the $\rho $
-parameter and to the muon decay
constant in a two-doublet model. Analyses along the
line of Peskin and Takeuchi
[8] are also available [9-10].
All of these analyses are, however, restricted to the
oblique-type corrections. It is now apparent that other
processes should be investigated in the two-doublet model
as much as possible from the view point of the internal
loop effects of Higgs boson masses.  The two-doublet
model could allow for large radiative corrections and
would provide us with useful information.

On the experimental side, it has been considered that
longitudinally polarized gauge boson scatterings,  i.e.,
$W_{L}^{+}W_{L}^{-}\rightarrow W_{L}^{+}W_{L}^{-}$,
$W_{L}^{+}W_{L}^{-}\rightarrow Z_{L}Z_{L}$,
are suitable processes to discover Higgs bosons in
future hadron colliders [11]. An elaborate calculations
of these reactions have been carried out in the
minimal standard model
(with a single Higgs  doublet)
including higher order radiative corrections [12-14].
The  decay process
$H ({\rm Higgs})\rightarrow W_{L}^{+}W_{L}^{-}$
constitutes sub-diagrams of the gauge boson scatterings
and the decay width has also been evaluated in the minimal
standard model.
(For experimental aspects of measuring the decay width,
see Ref. [15].)

Considering the importance of the $WW$-scattering and
the decay process, we have decided in the present
paper to undertake the calculation  of the
Higgs decay width  into a
$W_{L}^{+}W_{L}^{-}$   pair
in the  two-Higgs doublet model including loop corrections.
Our theoretical motivation is  to scrutinize the
dependence of the width on various Higgs boson masses
and explore the possibility of getting indirect signature
of  Higgs bosons other than $H$, which are yet to be
discovered.
There are two CP-even  neutral Higgs bosons ($H$, $h$ )
and a CP-odd one ($A$) together with  a charged one
($G^{\pm}$) in the  two Higgs doublet model.
In the present paper, we will assume that the $H$-boson is
the lightest among these scalar particles but is still
much heavier than the $W$-boson.  We will  study the
internal loop effects due to the other heavier scalar
particles, $h$, $A$, and $G^{\pm}$  to the decay process
$H \rightarrow W_{L}^{+}W_{L}^{-}$.

There have been several attempts to get
information on the
masses of these bosons.  According to the tree
unitarity  analysis [16] of the type of Lee,
Quigg and Thacker
[17], these masses are bounded from above as
$m_{H}<500 {\rm GeV}$,
$m_{h}<710 {\rm GeV}$,
$m_{G}<870 {\rm GeV}$,
$m_{A}<1200 {\rm GeV}$.
This is a criterion of the validity of perturbation
theory.   Similar bounds are also obtained on the
basis of triviality arguments [18].
Suppose that  the lightest Higgs boson $H$
has been discoverd in a future collider.   The
reasonable question to be raised thereby will be
whether one could glimpse into the existence of
another Higgs
boson in the mass range mentioned above
by looking at the decay width  of the discoverd
$H$.  In such a situation the calculation in the
present paper will become very useful.

This paper is organized as follows.
We describe the  two-Higgs model in Sec. 2 and
discuss the outline of our calculation.
We then proceed in Sec. 3 to calculate the tadpole
diagrams and various two-point functions.
In Sec. 4 we give the calculation of the Higgs
decay vertex function.   Numerical analyses
of the decay width formula are given in Sec. 5.
Sec. 6  is devoted to summary and
discussions.

\hskip1cm
\begin{center}
\large
{\bf 2. The Two-Higgs  Doublet Model}
\end{center}

Let us begin with the  Higgs potential,
consisiting of two Higgs doublets, $\Phi _{1}$ and
$\Phi _{2}$ with $Y=1$.
The most general $SU(2)_{L}\times U(1)_{Y}$ invariant
Higgs potential  becomes [19]
\begin{eqnarray}
V(\Phi _{1}, \Phi _{2})&=&
-\mu _{1}^{2}\mid \Phi _{1}\mid  ^{2}
-\mu _{2}^{2}\mid \Phi _{2}\mid ^{2}
-(\mu _{12}^{2}\Phi _{1}^{\dag}\Phi _{2}+\mu _{12}^{2*}
\Phi _{2}^{\dag}\Phi _{1})       \nonumber \\
& &
+\lambda _{1}\mid \Phi _{1}\mid ^{4}
+\lambda _{2}\mid \Phi _{2}
\mid ^{4}
+ \lambda _{3}\mid \Phi _{1}\mid  ^{2}\mid  \Phi _{2}\mid
^{2}  \nonumber \\
& &+ \lambda _{4}({\rm Re}\Phi _{1}^{\dag}\Phi _{2})^{2}
+\lambda _{5}({\rm Im }\Phi _{1}^{\dag}\Phi _{2})^{2}.
\end{eqnarray}
It is imperative  to avoid the flavor changing
neutral current and  henceforth the discrete
symmetry under $\Phi _{2} \rightarrow -\Phi _{2}$
[20] has been assumed except for the soft breaking term
$
(\mu _{12}^{2}\Phi _{1}^{\dag}\Phi _{2}+\mu _{12}^{2*}
\Phi _{2}^{\dag}\Phi _{1})
$.
This soft breaking term may be important in connection
 with a new
source of CP-violation for baryon genesis [21].
 Inclusion of this term, however, will render our
 calculation clumsy to some extent,
 because of the mixing of CP-even  and  odd states [22].
 Only for this this reason, we set
$\mu _{12}=\mu _{12}^{*}=0$
throughout the present paper. Inclusion of mixing
between  CP-even and odd  states will be discussed in
our future publications.

In order to see the particle contents and mass spectrum,
we rewrite  the Higgs potential in terms of
 the parametrization of the Higgs  doublets
\begin{equation}
\Phi _{i}=\left (
\begin{array}{cc}
w_{i}^{\dag}   \\
\frac{1}{\sqrt{2}}(v_{i}+h_{i}+iz_{i})
\end{array}
\right ),
\end{equation}
where  the vacuum expectation values $v_{1}$ and $v_{2}$
triggering the spontaneous break down,  are assumed to
be positive without spoiling generality.

The mass term in Eq. (1) may be diagonalized by introducing
 two kinds of mixing angles, $\alpha$ and $\beta $
  in the following way
\begin{equation}
\left (
\begin{array}{c}
h_{1} \\ h_{2}
\end{array}
\right )
=
\left (
\begin{array}{cc}
\cos \alpha & -\sin \alpha \\
\sin \alpha &  \cos \alpha \\
\end{array}
\right )
\left (
\begin{array}{c}
h \\ H
\end{array}
\right ),
\end{equation}

\begin{equation}
\left (
\begin{array}{c}
w_{1} \\ w_{2}
\end{array}
\right )
=
\left (
\begin{array}{cc}
\cos \beta & -\sin \beta  \\
\sin \beta &  \cos \beta   \\
\end{array}
\right )
\left (
\begin{array}{c}
w \\ G
\end{array}
\right ),
\end{equation}
\begin{equation}
\left (
\begin{array}{c}
z_{1} \\ z_{2}
\end{array}
\right )
=
\left (
\begin{array}{cc}
\cos \beta & -\sin \beta  \\
\sin \beta &  \cos \beta   \\
\end{array}
\right )
\left (
\begin{array}{c}
z \\ A
\end{array}
\right ).
\end{equation}
The mixing angles are determined in fact by the following
relations.
\begin{equation}
\tan 2\alpha =\frac{(\lambda _{3}+\lambda _{4})
v_{1}v_{2}}{
\lambda _{1}v_{1}^{2}-\lambda _{2}v_{2}^{2}}
, \hskip1cm
  \tan \beta=\frac{v_{2}}{v_{1}}
\end{equation}
($\pi /2 \geq \alpha \geq -\pi /2, \pi /2 > \beta
 > 0$).
The Nambu-Goldstone bosons to be absorbed into the
longitudinal part of $W^{\pm }$ and $Z$ are denoted by
$w$ and $z$ respectively.   Other  fields $h$, and $H$
are neutral while $G$ is a charged one.
The five coupling constants  in Eq. (1) are expressed by
the masses of these scalar particles together with the
mixing angles [23]

\begin{equation}
\lambda _{1}=\frac{1}{2v^{2}\cos ^{2}\beta}(m_{h}^{2}
\cos ^{2}\alpha +m_{H}^{2}\sin ^{2}\alpha ),
\end{equation}

\begin{equation}
\lambda _{2}=\frac{1}{2v^{2}\sin ^{2}\beta}(m_{h}^{2}
\sin ^{2}\alpha +m_{H}^{2}\cos ^{2}\alpha ),
\end{equation}

\begin{equation}
\lambda _{3}=\frac{\sin 2\alpha}{v^{2}\sin 2\beta}
(m_{h}^{2}-m_{H}^{2})+\frac{2m_{G}^{2}}{v^{2}},
\end{equation}

\begin{equation}
\lambda _{4}=-\frac{2m_{G}^{2}}{v^{2}},
\end{equation}

\begin{equation}
\lambda _{5}=\frac{2}{v^{2}}(m_{A}^{2}-m_{G}^{2}),
\end{equation}
where  $v=\sqrt{ v_{1}^{2}+v_{2}^{2}}\approx
246 {\rm GeV}$.
Incidentally,  masses $\mu _{1}^{2}$  and $\mu
_{2}^{2}$ in Eq. (1) are given on the tree level by
the vacuum stability condition
\begin{equation}
\mu _{1}^{2}=\lambda _{1}v_{1}^{2}+\frac{1}{2}(\lambda
 _{3}+\lambda _{4})v_{2}^{2},
\hskip1cm
\mu _{2}^{2}=\lambda _{2}v_{2}^{2}+\frac{1}{2}(\lambda
 _{3}+\lambda _{4})v_{1}^{2}.
\end{equation}
Eqs. (7)-(12) define the change of our initial
 seven parameters
 ($\lambda _{i} (i=1-5), \mu _{1}^{2}, \mu _{2}^{2})$
 into the set
($m_{h}$, $m_{H}$, $m_{G}$, $m_{A}$, $\alpha$,
$\beta$, $v$).

We are now interested in the decay process $H\rightarrow
W_{L}^{+}+W_{L}^{-}$.  We will use the equivalence
theorem [23, 17] which states
that if $m_{H}^{2} \gg M_{W}^{2}$,
we are allowed to replace the external longitudinal gauge
bosons by the corresponding Nambu-Goldstone bosons.
In our case, therefore, the evaluation of the process
 $H\rightarrow w^{+}w^{-} $ is our central  concern.
The tree-level  interaction term dictating this process is
extracted from the potential (1) as
\begin{equation}
{\cal L}_{Hww}=\frac{m_{H}^{2}}{v}\sin (\alpha
-\beta)Hw^{\dag}w.
\end{equation}
One might suspect that the counter terms corresponding
to this part of the potential  are just obtained by varying
the coefficients in (13), that is, by putting
$m_{H}^{2}\rightarrow m_{H}^{2}-\delta m_{H}^{2}$,
$v\rightarrow v-\delta v$,
$\alpha \rightarrow \alpha -\delta \alpha$,
and
$\beta \rightarrow \beta  -\delta \beta  $,
together with the wave
function renormalization.
This, however, does not give us all the counter terms.
There are other terms coming from the state mixing.
Changes of mixing angles $\delta
\alpha $ and $\delta \beta$ induce a mixture of the
paired fields, $H\leftrightarrow h$, and
$w\leftrightarrow G$, respectively.  This indicates that
some of the counterterms are obtained from the
$hw^{\dag}w$,  $HGw^{\dag}$, and $HG^{\dag}w$ vertices
\begin{equation}
{\cal L}_{hww}=-\frac{m_{h}^{2}}{v}\cos (\alpha
-\beta)hw^{\dag}w,
\end{equation}
\begin{equation}
{\cal L}_{HGw}=-\frac{m_{H}^{2}-m_{G}^{2}}{v}\cos
 (\alpha -\beta)(Hw^{\dag}G+HwG^{\dag}),
\end{equation}
by replacing $h\rightarrow - \delta \alpha H$ and $G
\rightarrow  \delta \beta w$.

After all we conclude that the counter terms
 for the $Hw^{\dag}w$ vertex take the following form
\begin{eqnarray}
\delta {\cal L}_{Hww}&=& \{
-\frac{\delta m_{H}^{2}}
{m_{H}^{2}}+\frac{\delta v}{v}
+(\sqrt{Z_{H}}-1)+(Z_{w}-1)\}{\cal L}_{Hww}
            \nonumber \\
& &+\delta \alpha \frac{m_{h}^{2}-m_{H}^{2}}
{v}\cos (\alpha
-\beta)Hw^{\dag }w
\nonumber \\
& &+\delta \beta\frac{2m_{G}^{2}-m_{H}^{2}}{v}
\cos (\alpha -\beta)Hw^{\dag}w.
\nonumber \\
\end{eqnarray}
Thus to evaluate the radiative corrections
to this process, we have to know $\delta m_{H}^{2}$,
$\delta v$, $\delta \alpha $,
 $\delta \beta$,
together with the wave function renormalization constants
 $Z_{w}$ and $Z_{H}$.

The calculation of $\delta v /v$ is fascilitated by
considering the renormalization of the $W$-boson mass
$M_{W}$, {\it i.e.}
$\delta v/v=\delta M_{W}^{2}/2M_{W}^{2}$.
This has been computed previously by Toussaint [5]
 and we just quote his  results,
\begin{eqnarray}
\frac{\delta M_{W}^{2}}{M_{W}^{2}}&=&
\frac{1}{(4\pi )^{2}v^{2}}\{\frac{1}{2}(2m_{G}
^{2}+m_{H}^{2}+m_{h}^{2}+
m_{A}^{2})+\frac{m_{G}^{2}m_{A}^{2}}
{m_{A} ^{2}-m_{G}^{2}}\ln\frac{m_{G}^{2}}
{m_{A}^{2}}        \nonumber \\
& &+\cos ^{2}(\alpha -\beta)\frac{m_{G}^{2}m_{H}^{2}}
{m_{H}^{2}-m_{G}^{2}}\ln\frac{m_{G}^{2}}{m_{H}^{2}}
+\sin ^{2}(\alpha -\beta)\frac{m_{G}^{2}m_{h}^{2}}
{m_{h}^{2}-m_{G}^{2}}\ln \frac{m_{G}^{2}}{m_{h}^{2}}
\}.                  \nonumber \\
\end{eqnarray}
Note that the logarithmic terms are all negative.
Since $m_{h}^{2}>m_{H}^{2}$, $\delta M_{W}^{2}$
is minimized at $\sin ^{2}(\alpha -\beta)=1$.

\hskip1cm
\pagebreak
\begin{center}
\large
{\bf 3. Tadpole Diagrams and Self-Energies}
\end{center}

Before launching into the computation of the radiative
corrections to $Hw^{\dag }w$ vertex, we prepare  the
counter terms (16) by evaluating $\delta m_{H}^{2}$,
$\delta \alpha $, $\delta \beta $,
$Z_{w}$, and $Z_{H}$.  First of all,
the condition of the stability of the vacuum (12) must be
replaced on the loop-level by including the tadpole
diagrams in Fig. 1.  The vacuum expectation values $v_{1}$
and $v_{2}$ are determined in our case by
\begin{equation}
\mu _{1}^{2}=\lambda _{1}v_{1}^{2}+\frac{1}{2}(\lambda
_{3}+\lambda _{4})v_{2}^{2}+T_{H}\frac{\sin \alpha}
{v_{1}}-T_{h}\frac{\cos \alpha }{v_{1}},
\end{equation}
\begin{equation}
\mu _{2}^{2}=\lambda _{2}v_{2}^{2}+\frac{1}{2}(\lambda
_{3}+\lambda _{4})v_{1}^{2}-T_{H}\frac{\cos \alpha}
{v_{2}}-T_{h}\frac{\sin \alpha }{v_{2}}.
\end{equation}
Here tadpole contributions of Fig. 1 are given by
\begin{eqnarray}
T_{H}&=&\frac{3m_{H}^2}{2v}
(\frac{\sin ^{3}\alpha }{\cos \beta} -\frac{\cos ^{3}
\alpha }{\sin \beta})f_{1}(m_{H}^{2})     \nonumber \\
& &-\frac{1}{v}(m_{h}^{2}+\frac{1}{2}m_{H}^{2})
\frac{\sin 2\alpha}{\sin 2\beta}\sin(\alpha -\beta )
f_{1}(m_{h}^{2})  \nonumber                         \\
& &+\{\frac{m_{H}^{2}}{2v}(\frac{\sin \alpha \sin ^{2}
\beta}{\cos \beta}-\frac{\cos \alpha \cos ^{2}\beta}
{\sin \beta})+\frac{m_{A}^{2}}{v}\sin (\alpha -
\beta)\}f_{1}(m_{A}^{2})    \nonumber        \\
& &+\{\frac{m_{H}^{2}}{v}(\frac{\sin \alpha \sin ^{2}
\beta}{\cos \beta}-\frac{\cos \alpha \cos ^{2}\beta}
{\sin \beta})+\frac{2m_{G}^{2}}{v}\sin (\alpha -
\beta)\}f_{1}(m_{G}^{2}),      \nonumber  \\
\end{eqnarray}
\begin{eqnarray}
T_{h}&=&-\frac{3m_{h}^2}{2v}
(\frac{\cos ^{3}\alpha }{\cos \beta} +\frac{\sin ^{3}
\alpha }{\sin \beta})f_{1}(m_{h}^{2})     \nonumber \\
& &-\frac{1}{v}(m_{H}^{2}+\frac{1}{2}m_{h}^{2})
\frac{\sin 2\alpha}{\sin 2\beta}\cos(\alpha -\beta )
f_{1}(m_{H}^{2})  \nonumber                         \\
& &-\{\frac{m_{h}^{2}}{2v}(\frac{\cos \alpha \sin ^{2}
\beta}{\cos \beta}+\frac{\sin \alpha \cos ^{2}\beta}
{\sin \beta})+\frac{m_{A}^{2}}{v}\cos (\alpha -
\beta)\}f_{1}(m_{A}^{2})    \nonumber        \\
& &-\{\frac{m_{h}^{2}}{v}(\frac{\cos \alpha \sin ^{2}
\beta}{\cos \beta}+\frac{\sin \alpha \cos ^{2}\beta}
{\sin \beta})+\frac{2m_{G}^{2}}{v}\cos (\alpha -
\beta)\}f_{1}(m_{G}^{2}).      \nonumber  \\
\end{eqnarray}
The definition of the functin $f_{1}(m^{2})$ is
given  in  Appendix A.

Let us next turn to the self-energy diagrams of $w$.
 The two-point function depicted in Fig. 2
becomes $2\times 2$   matrix due to the mixing with
charged  Higgs boson  $G$.
The two-point functions are decomposed into three terms
 according to the three diagrams in Figs 2(a), 2(b)
and 2(c), respectively,
\begin{equation}
\Pi _{ww}(p^{2})=\Pi _{ww}^{(1)}+\Pi _{ww}^{(2)}+
\Pi _{ww}^{(3)},
\end{equation}
\begin{equation}
\Pi _{wG}(p^{2})=\Pi _{wG}^{(1)}+\Pi _{wG}^{(2)}+
\Pi _{wG}^{(3)}.
\end{equation}
Internal particles in Figs. 2(a) and 2(c) are
tabulated in Table 1.
Our straightforward calculation
shows that the sum of the three terms are
summarized in the following form
\begin{eqnarray}
\Pi _{ww}(p^{2})&=&{\hat \Pi }_{ww}(p^{2})
-{\hat \Pi }_{ww}(0),              \\
\Pi _{wG}(p^{2})&=&{\hat \Pi }_{wG}(p^{2})
-{\hat \Pi }_{wG}(0).              \\
\nonumber
\end{eqnarray}
The explicit forms of ${\hat \Pi} _{ww}(p^{2})$ and
${\hat \Pi}_{wG}(p^{2})$ are given in Appendix B. Eqs.
 (24) and  (25) show  the existence of a massless pole
corresponding to the Nambu-Goldstone boson.  For a
single Higgs case it has been known that the calculation
of the self-energy may be made easier
by a manipulation
concceived  by Taylor [25].  In our case,
however,  we did
not develop similar tricks and just simply added up all
Feynman diagrams to reach Eqs. (24) and (25).

The wave function renormalization of the Nambu-Goldstone
Boson is finite and turns out to be, for the on-shell
renormalization,
\begin{eqnarray}
Z_{w}&=&1-\frac{1}{(4\pi )^{2}v^{2}}\{\frac{1}{2}(2m_{G}
^{2}+m_{H}^{2}+m_{h}^{2}+
m_{A}^{2})+\frac{m_{G}^{2}m_{A}^{2}}
{m_{A} ^{2}-m_{G}^{2}}\ln\frac{m_{G}^{2}}
{m_{A}^{2}}        \nonumber \\
& &+\cos ^{2}(\alpha -\beta)\frac{m_{G}^{2}m_{H}^{2}}
{m_{H}^{2}-m_{G}^{2}}\ln\frac{m_{G}^{2}}{m_{H}^{2}}
+\sin ^{2}(\alpha -\beta)\frac{m_{G}^{2}m_{h}^{2}}
{m_{h}^{2}-m_{G}^{2}}\ln \frac{m_{G}^{2}}{m_{h}^{2}}
\}.                  \nonumber \\
\end{eqnarray}

The renormalization of the mixing angle $\beta$
 is peculiar if we set the renormalization condition at
$p^{2}=0$. (A different renormalization scheme has been
used in Ref. [26].)  The vanishing of the off-diagonal part
  $\Pi _{wG}(p^{2})$ at $p^{2}=0$ tells us that there
is no corrections to the mixing angle $\beta$. In other
words the relation $\tan \beta =v_{2}/v_{1}$ is
preserved to higher orders

\begin{equation}
\delta \beta =-\frac{1}{m_{G}^{2}}\Pi _{wG}(0)=0.
\end{equation}
The last term in Eq. (16) thus gives no contribution
 to our calculation.

The other renormalized quantities w.r.t.
the $H$ field are given similarly in terms of
the two-point functions,

\begin{equation}
Z_{H}=1+\Pi _{HH}'(m_{H}^{2}),
\end{equation}
\begin{equation}
\delta m_{H}^{2}={\rm Re}(\Pi _{HH}(m_{H}^{2})),
\end{equation}
\begin{equation}
\delta \alpha =\frac{1}{m_{h}^{2}-m_{H}^{2}}
{\rm Re }(\Pi _{hH}(m_{H}^{2}) ).
\end{equation}

It is therefore an urgent task to evaluate
$\Pi _{HH}(p^{2})$ and $\Pi _{hH}(p^{2})$.
The relevant Feynman diagrams are again
 those in Fig. 2
 whose internal lines are explained in Table 2.
All of the results for these self-energies
are given in Appendices C and D.

The mixing phenomena between $H$ and $h$ may
be understood
in the following way. The $H$ quanta contain the
$h$ component due to the mixing $\Pi _{hH}(p^{2})$.
 Therefore
 $H$  quanta are able to transform  into $h$ and then
couple to $w^{\dag}w$  pair with the strength
determined by (14).
Such mixing phenomena are already
included in (16) through  the
the $\delta \alpha $ term (times $m_{h}^{2}\cos
(\alpha -\beta)$).

\hskip1cm
\begin{center}
\large
{\bf 4. Loop Corrections to the Higgs Vertex}
\end{center}

We are now in a position to evaluate the loop effects
to the $Hw^{\dag}w $ vertex as  shown in Fig. 3.
Hereafter the Nambu-Goldstone bosons are put on the
mass shell ($p_{1}^{2}=p_{2}^{2}=0$), but considering
future applications to
$W_{L}^{+}W_{L}^{-}\rightarrow
 W_{L}^{+}W_{L}^{-}$
we will keep the momentum $p$ carried by $H$ boson
off-shell and will
eventually put on the mass-shell
$(p^{2}=m_{H}^{2}$)  to get the decay
rate.

There are a lot of Feynman
diagrams to be calculated. We would like to
classify the diagrams into three groups
accoding to the three types in Fig. 4, that is,

\begin{equation}
\Gamma (p^{2})=\Gamma ^{(1)}+\Gamma ^{(2)}+
\Gamma ^{(3)}.
\end{equation}
These three terms come from  Figs. 4(a),  4(b)
 and 4(c), respectively.
 The species of the internal lines in these diagrams
are summarized in Table 3.

The sum of all the diagrams in Fig. 4 is rather lengthy,
 but is given here for our use in numerical
calculations.
As to the contribution of Fig. 4(a), we obtain
\begin{eqnarray}
\Gamma ^{(1)}&=&
-\frac{m_{H}^{6}}{v^{3}}\sin ^{3}(
\alpha -\beta)g(p^{2}, 0, 0, m_{H}^{2})
                                    \nonumber \\
& &-\frac{m_{h}^{4}m_{H}^{2}}{v^{3}}\sin
(\alpha -\beta)\cos ^{2}(\alpha -\beta) g(p^{2}, 0, 0,
 m_{h}^{2})    \nonumber \\
& &+\frac{1}{v^{3}}\{m_{H}^{2}(\frac{\cos \alpha
\cos ^{2}\beta
}{\sin \beta}-\frac{\sin \alpha \sin ^{2}\beta
}{\cos \beta})   -2m_{G}^{2}\sin (\alpha -\beta)\}
    \nonumber  \\
& &\times \{(m_{h}^{2}-m_{G}^{2})^{2}\sin ^{2}
(\alpha -\beta)g(p^{2}, m_{G}^{2},  m_{G}^{2},
 m_{h}^{2})
   \nonumber \\
& &+(m_{H}^{2}-m_{G}^{2})^{2}\cos ^{2}(\alpha -\beta)
g(p^{2},  m_{G}^{2},  m_{G}^{2},  m_{H}^{2})
     \nonumber \\
& &+(m_{A}^{2}-m_{G}^{2})^{2}g(p^{2},  m_{G}^{2},
m_{G}^{2},  m_{A}^{2})\}
  \nonumber \\
& &+\frac{1}{2v^{3}}\{m_{H}^{2}(\frac{\cos ^{2}
\beta \cos \alpha}{\sin \beta}-\frac{\sin ^{2}
\beta \sin \alpha}{\cos \beta})-2m_{A}^{2}
\sin (\alpha -\beta)\}
    \nonumber \\
& &\times (m_{A}^{2}-m_{G}^{2})^{2}g(p^{2},
  m_{A}^{2},  m_{A}^{2},  m_{G}^{2})
                  \nonumber \\
& &+\frac{m_{h}^{2}(m_{H}^{2}-m_{G}^{2})(m_{h}^
{2}-m_{G}^{2})}{v^{3}}\sin (\alpha -\beta)
\cos ^{2}(\alpha -\beta)
g(p^{2},  0,  m_{G}^{2},  m_{h}^{2})
   \nonumber \\
& &-\frac{m_{H}^{2}(m_{H}^{2}-m_{G}^{2})^{2}
}{v^{3}}\sin (\alpha -\beta)
\cos ^{2}(\alpha -\beta)
g(p^{2},  0,  m_{G}^{2},  m_{H}^{2})
   \nonumber \\
& &+\frac{3m_{H}^{6}}{v^{3}}\sin ^{2}(\alpha -
\beta)(\frac{\cos ^{3}\alpha }{\sin \beta}-
\frac{\sin ^{3}\alpha}{\cos \beta})g(p^{2},
  m_{H}^{2},  m_{H}^{2},  0)
       \nonumber \\
& &+\frac{3m_{H}^{2}(m_{H}^{2}-m_{G}^{2})
^{2}}{v^{3}}\cos ^{2}(\alpha -
\beta)(\frac{\cos ^{3}\alpha }{\sin \beta}-
\frac{\sin ^{3}\alpha}{\cos \beta})g(p^{2},
  m_{H}^{2},  m_{H}^{2},  m_{G}^{2})
       \nonumber \\
& &-\frac{m_{h}^{2}m_{H}^{2}(2m_{H}^
{2}+m_{h}^{2})}{v^{3}}
\sin (\alpha -\beta)\cos ^{2}(\alpha -\beta)
\frac{\sin 2\alpha}{\sin 2\beta}
g(p^{2},  m_{H}^{2},  m_{h}^{2},  0)
      \nonumber \\
& &+\frac{(m_{h}^{2}-m_{G}^{2})(m_{H}^{2}
-m_{G}^{2})(2m_{H}^{2}+m_{h}^{2})}{v^{3}}
\sin (\alpha -\beta)\cos ^{2}(\alpha -\beta)
\frac{\sin 2\alpha}{\sin 2\beta}
       \nonumber \\
& &\times g(p^{2},  m_{H}^{2},  m_{h}^{2},  m_{G}^{2})
      \nonumber \\
& &+\frac{m_{h}^{4}(m_{H}^{2}+2m_{h}^{2})}{v^{3}}
\sin (\alpha -\beta)\cos ^{2}(\alpha -\beta)
\frac{\sin 2\alpha}{\sin 2\beta}
g(p^{2},  m_{h}^{2},  m_{h}^{2},  0)
                           \nonumber \\
& &+\frac{(m_{h}^{2}-m_{G}^{2})^{2}
(m_{H}^{2}+2m_{h}^{2})}{v^{3}}
\sin ^{3}(\alpha -\beta)
\frac{\sin 2\alpha}{\sin 2\beta}
g(p^{2},  m_{h}^{2},  m_{h}^{2},  m_{G}^{2}).
                           \nonumber \\
\end{eqnarray}
The Feynman integral
is expressed in terms of
the function $g(p^{2}, m_{1}^{2}, m_{2}^{2}, m_{3}^{2})$
defined in Appendix A.
 As is shown there, this function
 is a combionation of the so-called Spence function.
(See Ref. [27] for a concise
review of the Spence function.)
It is almost straightforward to evaluate $\Gamma ^{(1)}$
 numerically on computer.

The diagrams in Fig. 4(b) are expressed by
the function
$f_{2}(p^{2}, m_{1}^{2}, m_{2}^{2})$
in defined in Eq. (38)  and provide us with
\begin{eqnarray}
\Gamma ^{(2)}&=&
\frac{5m_{H}^{2}}{2v^{3}}\{m_{h}^{2}\cos ^{2}(\alpha
-\beta)+m_{H}^{2}\sin ^{2}(\alpha -\beta)\}
\sin (\alpha -\beta)f_{2}(p^{2},  0,  0)
         \nonumber \\
& &-\frac{1}{v^{3}}\{m_{H}^{2}(\frac{\cos \alpha\cos
^{2}\beta}{\sin \beta}-\frac{\sin \alpha \sin ^{2}
\beta}{\cos \beta})-2m_{G}^{2}\sin (\alpha -\beta)
\}        \nonumber \\
& &\times \{
2m_{h}^{2}\sin ^{2}(\alpha -\beta)+2m_{H}^{2}
\cos ^{2}(\alpha -\beta)
           \nonumber \\
& &+\frac{\sin 2\alpha}{\sin 2\beta}
(m_{h}^{2}-m_{H}^{2})+m_{A}^{2}\}f_{2}(p^{2},
m_{G}^{2},  m_{G}^{2})  \nonumber \\
& &-\frac{1}{v^{3}}\{m_{H}^{2}(\frac{\cos \alpha\cos
^{2}\beta}{\sin \beta}-\frac{\sin \alpha \sin ^{2}
\beta}{\cos \beta})-2m_{A}^{2}\sin (\alpha -\beta)
\}        \nonumber \\
& &\times \{
\frac{1}{2}m_{h}^{2}\sin ^{2}(\alpha -\beta)+\frac
{1}{2}m_{H}^{2}\cos ^{2}(\alpha -\beta)
           \nonumber \\
& &+\frac{1}{2}\frac{\sin 2\alpha}{\sin 2\beta}
(m_{h}^{2}-m_{H}^{2})+m_{G}^{2}\}f_{2}(p^{2},
m_{A}^{2},  m_{A}^{2})   \nonumber \\
& &-\frac{1}{v^{3}}(m_{h}^{2}-m_{H}^{2})
(m_{H}^{2}-m_{A}^{2})\sin (\alpha -\beta)
\cos ^{2}(\alpha -\beta)f_{2}(p^{2},  m_{A}^{2},
  0)
             \nonumber \\
& &-\frac{4}{v^{3}}(m_{h}^{2}-m_{H}^{2})
(m_{H}^{2}-m_{G}^{2})\sin (\alpha -\beta)
\cos ^{2}(\alpha -\beta)f_{2}(p^{2},  m_{G} ^{2},
  0)
             \nonumber \\
& &-\frac{3m_{H}^{2}}{2v^{3}}
\{
((m_{h}^{2}-m_{H}^{2})\frac{\sin 2\alpha}{\sin 2\beta}
+2m_{G}^{2})\cos ^{2}(\alpha -\beta)
          \nonumber \\
& &+m_{H}^{2}\}
(\frac{\cos ^{3}\alpha}
{\sin \beta}-\frac{\sin ^{3}\alpha}{\cos \beta})
f_{2}(p^{2},  m_{H}^{2},  m_{H}^{2})
             \nonumber \\
& &-\frac{1}{v^{3}}\{(m_{h}^{2}-m_{H}^{2})\frac{
\sin 2\alpha}{\sin 2\beta}+2m_{G}^{2}\}(m_{h}^{2}
+2m_{H}^{2})
              \nonumber \\
& &\times \sin (\alpha -\beta)\cos ^{2}(\alpha -
\beta)\frac{\sin 2\alpha }{\sin 2\beta}
f_{2}(p^{2},  m_{H}^{2},   m_{h}^{2})
        \nonumber \\
& &-\frac{1}{2v^{3}}\{((m_{h}^{2}-m_{H}^{2})\frac{
\sin 2\alpha}{\sin 2\beta}+2m_{G}^{2})\sin ^{2}
(\alpha -\beta)+m_{h}^{2}
\}            \nonumber \\
& &\times (m_{H}^{2}+2m_{h}^{2})\sin (\alpha -\beta )
\frac{\sin 2\alpha }{\sin 2\beta}f_{2}(p^{2},
m_{h}^{2},   m_{h}^{2}).
              \nonumber \\
\end{eqnarray}

Finally we give the results of Fig. 4(c)
which are multiplied by 2,  because we have
to include symmetric diagrams as well
\begin{eqnarray}
\Gamma ^{(3)}&=&
-\frac{2m_{h}^{2}}{v^{3}}\{(m_{h}^{2}-
m_{H}^{2})\frac{\sin 2\alpha}{\sin 2\beta}
+2m_{G}^{2}\}\sin (\alpha -\beta )
\cos ^{2}(\alpha -\beta )f_{2}(0, m_{h}^{2}, 0)
              \nonumber \\
& &+\frac{2m_{H}^{2}}{v^{3}}\{(m_{h}^{2}
\frac{\sin 2\alpha }{\sin 2\beta}+2m_{G}^{2})
\sin (\alpha -\beta)\cos ^{2}(\alpha -\beta)
            \nonumber  \\
& & +m_{H}^{2}(\frac{\sin ^{3}\alpha }{\cos \beta}
-\frac{\cos ^{3}\alpha}{\sin \beta})\sin ^{2}(\alpha
-\beta)\}f_{2}(0, m_{H}^{2}, 0)
      \nonumber \\
& &-\frac{2(m_{h}^{2}-m_{G}^{2})}{v^{3}}
\{\frac{\sin 2\alpha}{\sin 2\beta }(m_{h}^{2}
\sin ^{2}(\alpha -\beta )
+m_{H}^{2}\cos ^{2}(\alpha -\beta)
)
\nonumber \\
& &-m_{G}^{2}\cos (2\alpha
-2\beta)\}\sin (\alpha -\beta)f_{2}(0,
m_{G}^{2}, m_{h}^{2})
           \nonumber \\
& & -\frac{m_{H}^{2}-m_{G}^{2}}{v^{3}}\{
(m_{h}^{2}\frac{\sin 2\alpha}{\sin 2\beta}+
2m_{G}^{2})\sin (2\alpha -2\beta)
      \nonumber \\
& &+2m_{H}^{2}(\frac{\cos ^{3}\alpha}{\sin \beta}
-\frac{\sin ^{3}\alpha}{\cos \beta})
\cos (\alpha -\beta)\}
\cos (\alpha -\beta)f_{2}(0, m_{G}^{2}, m_{H}^{2})
     \nonumber \\
& &+2\frac{(m_{A}^{2}-m_{G}^{2})^{2}}{v^{3}}
\sin (\alpha -\beta)f_{2}(0, m_{G}^{2}, m_{A}^{2}).
\end{eqnarray}

Putting all the calculations together we arrive at
 the decay width formula
\begin{equation}
\Gamma (H\rightarrow W_{L}^{+}W_{L}^{-})=
\frac{1}{16\pi}\frac{1}{m_{H}}\sqrt{1-\frac{
4M_{W}^{2}}{m_{H}^{2}}}\vert {\cal M}(p^{2}=m_{H}^{2}
) \vert ^{2}.
\end{equation}
Here the invariant amplitude is given through the
one-loop
 order by the following sum
\begin{eqnarray}
{\cal M}(p^{2})&=& \Gamma (p^{2})+ \frac{1}{v}\cos
(\alpha -\beta)
\Pi _{hH}(m_{H}^{2})-\frac{1}{v}\sin (\alpha -\beta)
\Pi _{HH}(m_{H}^{2})    \nonumber \\
& &+\{\frac{\delta v}{v} +\frac{1}{2}\Pi _{HH}'(
m_{H}^{2})+Z_{w}\}\frac{m_{H}^{2}}{v}\sin (\alpha
-\beta).
\end{eqnarray}

Note that, while $\Gamma (p^{2})$,  $\Pi _{hH}
(m_{H}^{2})$,   and $\Pi _{HH}(m_{H}^{2})$
are all divergent, the combination
of these with the weight  in
Eq. (36) is  finite.  Although this is guaranteed by
the renormalizability, we have confirmed the
finiteness explicitly by hand.  This is a non-trivial
check of our calculation.

Suppose that $m_{G}$ is much larger than
 all the other masses. It then turns out that
 $\Gamma (p^{2})$  will be dominated
 ( on the dimensional account) by the
 terms of the  form $m_{G}^{4}/v^{3}$ with
 possible logarithmic corrections.  The same things
 also happen to the two-point functions
 $\Pi _{HH}(p^{2})$ and $\Pi _{hH}(p^{2})$,
 which will be
  principally given a form proportional
  to $m_{G}^{4}/v^{2}$.  Thus the decay width
is potentially very sensitive to the choice of
$m_{G}$, provided that $m_{G}$ is large.
This situation is in contrast to the standard model
, where the effect of the scalar boson is always veiled.
Of course the sensitivity to $m_{G}$ depends upon
 the values of $\alpha $ and $\beta $, and
it is interesting to see the behavior of the decay width
for various choices of $\alpha $ and $\beta $.

\hskip1cm
\begin{center}
\large
{\bf 5. Numerical Analysis of the Decay Width Formula}
\end{center}

Now let us  analyse the decay
width formulae (35) and (36)
 by putting numbers
 into  the parameters.  In principle we should keep
an open mind to look at every corner of the
parameter space.
There have been a lot of efforts to constrain the
parameters in the two-Higgs doublet model
 from phenomenological analyses,
and we will give due considerations
for an ecconomical purpose to reduce
the large parameter space.

Constarints on the mass of the charged Higgs
boson ($m_{G}$) and the mixing angle $\beta$ has been
discussed [28] by considering the low-energy data
relating to neutral meson mixing (
$K^{0}-{\bar K}^{0}$,
$D^{0}-{\bar D}^{0}$,
$B_{d}^{0}-{\bar B}_{d}^{0}$  ) and
CP-violation parameters.
It has been  argued that the low-energy data
exclude small $\tan \beta $ and light
$m_{G}$ (say, $\tan \beta <$ 0.3-0.4,
$m_{G}<200 $  GeV).  In the following
we will examine the decay width formula
by setting $\tan \beta =$ 2 and 10 as
tentative values.  These values are within a
perturvative region w.r.t. the ${\bar t}
bG$ coupling. We will also vary
the charged Higgs boson mass as 400 GeV $<
m_{G}<$ 900 GeV.  For the values in this
region it has  been known [6] that the $\rho$
parameter and the radiative corrections to the
muon decay constant do not contradict the
 present experimental values.
Since the $H$ boson is assumed to be the lightest,
  we fix $m_{H}=300$  GeV and also we take
$m_{h}=400$  GeV.
These values satisfy the tree unitarity constraint
as mentioned in Introduction.

As to the choice of the angle $\alpha $,
there does not seem to be a thorough
phenomenological analysis comparable to
the case of $\beta $.  Alternatively, the angle
$\alpha $ may be fixed in a theoretically
oriented way. In the minimal supersymmetric model
 for example,  the parameter $\alpha $ is not
independent but expressed in terms of  the Higgs
boson and gauge boson masses.  If one  takes the
charged Higgs boson mass as an independent parameter
and assumes an extremely large value for it, then
$\sin ^{2}(\alpha -\beta)$ approaches to unity.
In this case the strength of
 the $Hw^{\dag }w$ vertex (13)
becomes maximal.
In the following we will examine two extreme cases,
 namely,  $\sin ^{2}(\alpha -\beta)=1$ and
$\sin ^{2}( \alpha -\beta)=0$.
Note that, in the latter case,
the  $Hw^{\dag }w$ vertex vanishes on the tree level,
 and the decay process proceeds only due to loop
 effects.

To sum up we will evaluate the decay width (35)
as a function of $m_{G}$ for the following cases.
\begin{description}
\item{Case 1:}
\hskip0.5cm
$\tan \alpha =\tan \beta=2$,
 $m_{H}=300$ GeV,  $m_{h}=400$ GeV,
\item{Case 2:}
\hskip0.5cm
$\tan \alpha =\tan \beta=10$,
$m_{H}=300$ GeV,  $m_{h}=400$ GeV,
\item{Case 3:}
\hskip0.5cm
$\tan \beta=2$, $\sin ^{2}(\alpha -\beta )=1$,
$m_{H}=300$ GeV,  $m_{h}=400$ GeV,
\item{Case 4:}
\hskip0.5cm
$\tan \beta=10$,  $\sin ^{2}(\alpha -\beta )=1$,
$m_{H}=300$ GeV,  $m_{h}=400$ GeV,
\end{description}
As to the mass $m_{A}$, we will set tentatively
$m_{A}=350$ GeV,
$m_{A}=700$ GeV,  and
$m_{A}=1000$ GeV.
The result of our numerical computation  for the
cases 1, 2, 3 and 4 are illustrated in Figs.
 5, 6, 7 and 8,  respectively.

For the sake of comparison let us recall the decay
width formula in the one Higgs boson case.
The decay width in the minimal standard model (MSM) is
given through the one-loop order  by  [12]
\begin{eqnarray}
\Gamma ^{{\rm MSM}}(H\rightarrow W_{L}^{+}W_{L}^{-})&=&
\frac{1}{16\pi}\frac{m_{H}^{3}}{v^{2}}\sqrt{1-
\frac{4M_{W}^{2}}{m_{H}^{2}}}
\nonumber \\
& &\times \vert 1+\frac{1}{4\pi ^{2}}
\frac{m_{H}^{2}}{v^{2}}
(\frac{19}{16}-\frac{3\sqrt{3}\pi}{8}+
\frac{5\pi ^{2}}{48})\vert ^{2}.
\end{eqnarray}
This formula gives us
$\Gamma ^{{\rm MSM}}(H\rightarrow W_{L}^{+}W_{L}
^{-})\sim 7.6$ GeV for $m_{H}=300$ GeV.
Note that the radiative correction
in (37) is as small as 1.3 \%.  This is
due to the smallness of the coefficients
of $m_{H}^{2}/v^{2}$ in (37).

Let us now look at Figs. 5,6,7 and 8 closely.
We can immediately see  conspicuous differences
in the magnitude of computed  values in  these
four cases.  The cases 3 and 4 (Figs. 7 and 8) show
that the computed width ranges from 1 to 8
GeV, depending on $m_{G}$ and $m_{A}$.
In the case 1, on the other hand, the width is
 smaller by  one or two orders of magnitude
or perhaps more depending on $m_{G}$,
compared to the cases 3 and 4.
The case 2 also gives small decay width for
 400  GeV$<m_{G}<$ 700  GeV.

This situation may be  understood in the  following
way. As mentioned before, in the cases 1 and 2,
the tree
level coupling vanishes and the predicted decay
width
is  only due to the radiative corrections. The
radiative corrections thus turn out to be
small for the cases 1 and 2
for the same reason as in  the MSM case.
This is particularlly interesting,
if one would notice     that the values
$m_{G}=900$ GeV and $m_{A}=
1000$ GeV are on the verge of the breakdown of
perturvative calculation.

We also note that the $m_{G}$-dependence
of the width for cases 1
and 2 is  wild-behaved. It changes by one or two
orders of magnitude as we vary $m_{G}$ from 400 GeV
 to 900 GeV.  For large-$m_{G}$, the width goes up
 like $m_{G} ^{8}$, which is in accordance
with our initial expectation.
At any rate, if the width would be this  small as
in Fig. 5,   the experimental measurement of the width
would not be easy.

Since we have set $\sin ^{2}(\alpha -\beta )=1$
 for the cases 3 and 4, the tree-level prediction
of the width for these cases
coincide with the tree-level value in
the MSM case  without
radiative correction , which becomes
7.5 GeV.
In Figs. 7 and 8, we see that the radiative
correction becomes important for large $m_{G}$
for $m_{A}=350$ GeV and $m_{A}=700$ GeV.  The
corrections reaches 35 \% (18 \%) for $m_{A}=
350$ GeV ($m_{A}=700$ GeV) at $m_{G}=900$ GeV.
If the $m_{A}$ is as large as 1000 GeV, the radiative
corrections become comparable with the tree-level
 calculation for small $m_{G}$.  This is
attributed to the largeness of $m_{A}$ and shows
that the perturbative calculation is barely allowed
for this choice of $m_{A}$.
The $m_{G}$-dependence in Figs. 7 and 8 is
not so wild
as in Figs. 5 and 6.  This is due to the fact that the
wild $m_{G}^{4}/v^{3}$
behavior of the loop amplitudes is
tamed  by larger $m_{G}$-independent tree amplitudes.

We have repeated the above calculation by setting
$m_{h}=600$ GeV. The results, however, do not differ
much from those shown in Figs. 5,6,7 and 8.
This fact is also very interesting, since the
value $m_{h}=600 $  GeV is close to the boundary
of the perturbative region.  This insensitivity
to $m_{h}$ reminds us
of the analysis  by Bertolini [6].  He studied
radiative corrections, in the two-doublet model,
to the $\rho $-parameter and to the relation between
the Fermi constant and the $SU(2)_{L}$ gauge coupling
(often denoted by $\Delta r$).  He argued that,
for $\alpha =\beta$,  $\Delta r$ is independent of
$m_{h}$.

Suppose that the Higgs boson $H$ has been discovered
and the decay width $\Gamma (H\rightarrow W_{L}^{+}
W_{L}^{-}) $ has been measured to a good accuracy.
 In such a situation, the calculation shown in Figs.
7 and 8 may be useful to probe the existence of
another Higgs boson, {\it i.e.} G and A if we know
the values of $\alpha $ and $\beta$.
Of course we do not know at present
definite  value of  $\alpha $  or  $\beta$, either.
We have to look for various means of
measuring the parameters of the two-doublet model. Being
given various  measurements, we will be able to
fit the data
in a multi-dimensional parameter space.

Finally we add a comment of the $p^{2}$ dependence of
the invariant amplitude ${\cal M}(p^{2})$.  We have
examined ${\cal M}(p^{2})$  numerically in the hope
of possibility of future confrontation with
measurement.  With the
parameters of  cases  3 and 4, this form-factor
is almost a constant for 200 GeV $< \sqrt{p^{2}} <$
350 GeV, $m_{G}^{2}=500$ GeV, $m_{A}^{2}=$ 350 GeV,
700 GeV, and 1000 GeV,
and does not show a peculiar behavior.
This is because the $p^{2}$-independent tree amplitude
dominates over the loop amplitudes.
In the cases 1 and 2, on the other hand,
where the tree amplitude vanishes, the form factor goes
up rapidly as we increase $\sqrt{p^{2}}$  from 250 GeV
to 350 GeV.   It magnitude is, however, still very small
compared to the cases 3 and 4.

\hskip2cm
\begin{center}
\large
{\bf 6. Summary and Discussions}
\end{center}

In the present paper, we have investigated
the radiative corrections in the two-Higgs doublet
model from the viewpoint that
the loop corrections to the decay rate of
$H\rightarrow W_{L}^{+}W_{L}^{-}$
could be a useful probe into the other Higgs
bosons.  The corrections to the $Hw^{\dag}w$
 vertex contain terms behaving like
$m_{G}^{4}/v^{3}$ for large $m_{G}$.
This indicates that predicted width could be
 sensitive to the value $m_{G}$ .
We have demonstrated the behavior of the width
 as a function of $m_{G}$ for various choices
 of $m_{h}$, $m_{A}$, $\alpha $ and $\beta $.
We have seen to what extent the terms of the
power-behavior would become important in the
decay width formula.

The calculation presented in this paper can be
easily carried over to the decay $H \rightarrow
Z_{L}Z_{L}$, the analysis of which will be given
in our future publication [29].  The elastic
scattering $W_{L}^{+}W_{L}^{-}\rightarrow
W_{L}^{+}W_{L}^{-}$,  $Z_{L}Z_{L}$ is also of
particular interest. The present calculation
will become a useful  basis for the evaluation
of these elastic scattering.

The Higgs boson masses and the radiative
corrections in supersymmetric theories
have been one of important topics in recent
literatures [30].
The parameters in the Higgs sector in
supersymmetric theories are not completely
independent, but are constrained by the
supersymmetry.
Suppose that the decay rate of $H \rightarrow
W_{L}^{+}W_{L}^{-}$ be evaluated in supersymmetric
models.  The parameters describing the decay rate
can not be varied freely.
For such a case, it might be that the screening
phenomena could occur  effectively, because of the
constrained parameter space.
We will come to this problem in the future.

\hskip1cm
\begin{center}
\large
{\bf Acknowledgements}
\end{center}
One of the authors (T.K.) is grateful to
Professor C.S. Lim
for informative and helpful discussions.
The authors are grateful to Professor E. Takasugi
and Professor K. Higashijima for useful discussions.
This work is supported in part by the Grant in Aid
for Scientific Reserach from the Ministry of
Education, Science and Culture (no.  06640396).

\hskip1cm
\pagebreak
\begin{center}
\large
{\bf Appendix A}
\end{center}

Here we summarize our  notations for the Feynman
integrals  corresponding to various types of diagrams
in Fig. 2, which are expressed by
\begin{eqnarray}
f_{1}(m^{2})&=& \mu ^{4-D}
\int \frac{d^{D}k}{(2\pi )^{D}}\frac{i}
{k^{2}-m^{2}}   \nonumber \nonumber                 \\
&=&\frac{m^{2}}{(4\pi )^{2}}(\frac{2}{D-4}-1+
\gamma _{E}+\ln \frac{m^{2}}{4\pi \mu ^{2}}),
            \nonumber   \\
f_{2}(p^{2},m_{1}^{2},m_{2}^{2})&=&-i
\mu ^{4-D}\int \frac{
d^{D}k}{(2\pi )^{D}}\frac{i}{k^{2}-m_{1}^{2}}\frac{i}
{(p-k)^{2}-m_{2}^{2}}.   \nonumber \\
\end{eqnarray}
Here $\mu $ is the scale
parameter of the $D$-dimensional
regularization.
These notations will be used extensively in Appendices
B,C, and D.
If we set $m_{2}^{2}=0$ in Eq. (38), we obtain
\begin{eqnarray}
f_{2}(p^{2},m^{2},0)&=&
\frac{1}{(4\pi )^{2}}\{
\frac{2}{D-4} +\gamma _{E}-2+\ln \frac{m^{2}}{4\pi
\mu ^{2}}         \nonumber \\
& & +(1-\frac{m^{2}}{p^{2}})\ln (1-\frac{p^{2}}
{m^{2}})\}.   \nonumber \\
\end{eqnarray}
The formula (38) is also simplified if we put $m_{1}=
m_{2}=m$, i.e.,
\begin{eqnarray}
f_{2}(p^{2},m^{2},m^{2})&=&
\frac{1}{(4\pi )^{2}}\{
\frac{2}{D-4} +\gamma _{E}-2+\ln \frac{m^{2}}{4\pi
\mu ^{2}}       \nonumber \\
& &+\sqrt{1-\frac{4m^{2}}{p^{2}}}\ln \frac{
\sqrt{1-4m^{2}/p^{2}}+1}{\sqrt{1-4m^{2}/p^{2}}-1}\}.
   \nonumber \\
\end{eqnarray}

The vertex integral corresponding to Fig. 4
  is expressed
in terms of the following function
\begin{equation}
g(p^{2},  m_{1}^{2},  m_{2}^{2},  m_{3}^{2})=
\int \frac{d^{4}k}{(2\pi )^{4}}\frac{i}{(k-p_{1})^{2}-
m_{1}^{2}}\frac{i}{(k+p_{2})^{2}-m_{2}^{2}}
\frac{i}{k^{2}-m_{3}^{2}}.
\end{equation}
It has been known that this integral is given by
a combination of the Spence function;
\begin{eqnarray}
g(p^{2},  m_{1}^{2},  m_{2}^{2},  m_{3}^{2})&=&
\frac{1}{(4\pi )^{2}p^{2}}\ln (\frac{a-1}{a})\{\ln (a-\xi
 _{+})+\ln (a-\xi _{-})
       \nonumber \\
&+ &\ln (\frac{p^{2}}{(a+b)(m_{
 1}^{2}-m_{3}^{2})})\}
     \nonumber \\
&+&\frac{1}{(4\pi )^{2}p^{2}}\{
-{\rm Sp}(\frac{a-1}{a-\xi _{+}})
+{\rm Sp}(\frac{a}{a-\xi _{+}})
-{\rm Sp}(\frac{a-1}{a-\xi _{-}})
    \nonumber \\
&+&{\rm Sp}(\frac{a}{a-\xi _{-}})
+{\rm Sp}(\frac{a-1}{a+b})-{\rm Sp}(\frac{a}{a+b})\}.
\end{eqnarray}
Here the Spence function is defined as usual by
\begin{equation}
{\rm Sp}(z)=-\int _{0}^{z}\frac{dx}{x}\ln (1-x).
\end{equation}
Various quantities appearing in (42) are given by
\begin{equation}
\xi _{\pm}=\frac{m_{2}^{2}-m_{1}^{2}+p^{2}
\pm \sqrt{(m_{2}^{2}-m_{1}^{2}+p^{2})^{2}
-4m_{2}^{2}p^{2}}}{2p^{2}},
\end{equation}
\begin{equation}
a=\frac{m_{2}^{2}-m_{3}^{2}}{p^{2}},
\hskip1cm
b=\frac{m_{3}^{2}}{m_{1}^{2}-m_{3}^{2}}.
\end{equation}

\vskip1cm
\begin{center}
{\bf Appendix B}
\end{center}

The self-energy of the Nambu-Goldstone boson
$\Pi _{ww}(p^{2})$, and the two-point function
$\Pi _{wG}(p^{2})$  are
obtained by evaluating Fig. 2, whose internal
particles are specified in Table 1.
Straightforward calculations show that
these functions are  expressed as in (24)
and (25), where we have
introduced following functions
\begin{eqnarray}
{\hat \Pi }_{ww}(p^{2})&=&
\frac{m_{h}^{4}}{v^{2}}\cos ^{2}(
\alpha -\beta)f_{2}(p^{2},m_{h}^{2},0)
               \nonumber \\
& &+\frac{m_{H}^{4}}{v^{2}}\sin  ^{2}(
\alpha -\beta)f_{2}(p^{2},m_{H}^{2},0)
               \nonumber \\
& &+\frac{(m_{h}^{2}-m_{G}^{2})^{2}}{v^{2}}\sin ^{2}
(\alpha -\beta)f_{2}(p^{2},m_{h}^{2},m_{G}^{2})
               \nonumber \\
& &+\frac{(m_{H}^{2}-m_{G}^{2})^{2}}{v^{2}}\cos ^{2}
(\alpha -\beta)f_{2}(p^{2},m_{H}^{2},m_{G}^{2})
               \nonumber \\
& &+\frac{(m_{A}^{2}-m_{G}^{2})^{2}}{v^{2}}
f_{2}(p^{2},m_{A}^{2},m_{G}^{2}),
               \nonumber \\
\end{eqnarray}

\begin{eqnarray}
{\hat \Pi }_{wG}(p^{2})
&=&\frac{m_{h}^{2}(m_{h}^{2}-m_{G}^{2})}{2v^{2}}
\sin (2\alpha -2\beta)f_{2}(p^{2},
m_{h}^{2}, 0)     \nonumber \\
& &-\frac{m_{H}^{2}(m_{H}^{2}-m_{G}^{2})}{2v^{2}}
\sin (2\alpha -2\beta)f_{2}(p^{2},
m_{H}^{2}, 0)     \nonumber \\
& &+\frac{m_{h}^{2}-m_{G}^{2}}{v^{2}}\sin (\alpha -\beta)
\{m_{h}^{2}(\frac{\sin ^{2}\beta \cos \alpha }
{\cos \beta}+\frac{\cos ^{2}\beta \sin \alpha}
{\sin \beta})   \nonumber \\
& &+2m_{G}^{2}\cos (\alpha -\beta)\}
f_{2}(p^{2}, m_{h}^{2},m_{G}^{2})
                  \nonumber \\
& &+\frac{m_{H}^{2}-m_{G}^{2}}{v^{2}}\cos (\alpha -\beta)
\{m_{H}^{2}(\frac{\cos ^{2}\beta \cos \alpha }
{\sin \beta}-\frac{\sin ^{2}\beta \sin \alpha}
{\cos \beta})      \nonumber \\
& &-2m_{G}^{2}\sin (\alpha -\beta)\}
f_{2}(p^{2}, m_{H}^{2},m_{G}^{2}).
                  \nonumber \\
\end{eqnarray}

\begin{center}
{\bf Appendix C}
\end{center}

The self-energy of the $H$ boson is decomposed into
three parts
\begin{equation}
\Pi _{HH}(p^{2})= \Pi_{HH}^{(1)}+ \Pi
_{HH}^{(2)}+\Pi _{HH}^{(3)},
\end{equation}
which correspond to Figs. 2(a), 2(b) and 2(c),
 respectively.
These three diagrams give us the following
results:
\begin{eqnarray}
\Pi_{HH}^{(1)}&=&
\frac{9m_{H}^{4}}{2v^{2}}(\frac{\cos ^{3}\alpha }
{\sin \beta}-\frac{\sin ^{3}\alpha}{\cos \beta})^{2}
f_{2}(p^{2}, m_{H}^{2}, m_{H}^{2})  \nonumber \\
& &+\frac{(2m_{H}^{2}+m_{h}^{2})^{2}}{v^{2}}(\frac
{\sin 2\alpha }{\sin 2\beta })^{2}
\cos ^{2}(\alpha -\beta)f_{2}(p^{2}, m_{h}^{2}, m_{H}
^{2})        \nonumber   \\
& &+\frac{(2m_{h}^{2}+m_{H}^{2})^{2}}{2v^{2}}(\frac
{\sin 2\alpha }{\sin 2\beta })^{2}
\sin ^{2}(\alpha -\beta)f_{2}(p^{2},  m_{h}^{2},
m_{h}^{2}) \nonumber \\
& & +\frac{3m_{H}^{4}}{2v^{2}}\sin ^{2}(\alpha -\beta)
f_{2}(p^{2}, 0, 0)        \nonumber \\
& &+[\frac{m_{H}^{2}}{v}(\frac{\cos ^{2}\beta\cos
\alpha }{\sin \beta}-\frac{\sin ^{2}\beta \sin \alpha}
{\cos \beta})-\frac{2m_{G}^{2}}{v}\sin (\alpha -\beta)
]^{2}                   \nonumber \\
& & \times f_{2}(p^{2},  m_{G}^{2},  m_{G}^{2})
             \nonumber \\
& &+\frac{1}{2}[\frac{m_{H}^{2}}{v}
(\frac{\cos ^{2}\beta\cos
\alpha }{\sin \beta}-\frac{\sin ^{2}\beta \sin \alpha}
{\cos \beta})-\frac{2m_{A}^{2}}{v}\sin (\alpha -\beta)
]^{2}                   \nonumber \\
& & \times f_{2}(p^{2},  m_{A}^{2},  m_{A}^{2})
             \nonumber \\
& &+\frac{(m_{H}^{2}-m_{A}^{2})^2}{v^{2}}
\cos ^{2}(\alpha -\beta)f_{2}(p^{2},  m_{A}^{2},
0)          \nonumber \\
& &+\frac{2(m_{H}^{2}-m_{G}^{2})^2}{v^{2}}
\cos ^{2}(\alpha -\beta)f_{2}(p^{2},  m_{G}^{2},
0),          \nonumber \\
\end{eqnarray}

\begin{eqnarray}
\Pi _{HH}^{(2)}&=&
[\frac{m_{h}^{2}}{2v^{2}}(\frac{\cos \alpha \sin \beta }
{\cos ^{2}\beta}+\frac{\sin \alpha \cos \beta}
{\sin ^{2}\beta})\sin 2\alpha  \cos (\alpha
-\beta)                      \nonumber \\
& & +\frac{m_{H}^{2}}{v^{2}}
(\frac{\cos ^{3}\alpha}{\sin \beta}-\frac{
\sin ^{3}\alpha }{\cos \beta})
(\frac{\cos \alpha \cos ^{2}\beta}{\sin \beta}-
\frac{\sin \alpha \sin ^{2}\beta}{\cos \beta})
                \nonumber \\
& &+\frac{2m_{G}^{2}}{v^{2}}\sin ^{2}(\alpha -\beta)]
f_{1}(m_{G}^{2})               \nonumber \\
& &+[\frac{m_{h}^{2}}{4v^{2}}(\frac{\cos \alpha \sin
 \beta }
{\cos ^{2}\beta}+\frac{\sin \alpha \cos \beta}
{\sin ^{2}\beta})\sin 2\alpha  \cos (\alpha
-\beta)                      \nonumber \\
& & +\frac{m_{H}^{2}}{2v^{2}}
(\frac{\cos  ^{3}\alpha }{\sin \beta}-
\frac{\sin ^{3}\alpha }{\cos \beta})
(\frac{\cos \alpha \cos ^{2}\beta}{\sin \beta}-
\frac{\sin \alpha \sin ^{2}\beta}{\cos \beta})
           \nonumber \\
& &+\frac{m_{A}^{2}}{v^{2}}\sin ^{2}(\alpha -\beta)]
f_{1}(m_{A}^{2})               \nonumber \\
& &+[\frac{m_{h}^{2}}{2v^{2}}\{3(\frac{\sin 2\alpha }
{\sin 2\beta })^{2}
\sin ^{2}(\alpha -\beta)+\frac{\sin 2\alpha }
{\sin 2\beta }\}       \nonumber \\
& &+\frac{m_{H}^{2}}{2v^{2}}\{3(\frac{\sin 2\alpha }
{\sin 2\beta })^{2}\cos  ^{2}(\alpha -\beta)-
\frac{\sin 2\alpha }{\sin 2\beta }\}
]f_{1}(m_{h}^{2})
       \nonumber \\
& &+[\frac{3m_{h}^{2}}{2v^{2}}(\frac{\sin ^{2}\alpha
\cos \alpha}{\cos \beta}+\frac{\sin \alpha \cos ^{2}
\alpha}{\sin \beta})^{2}+
\frac{3m_{H}^{2}}{2v^{2}}(\frac{\sin ^{3}\alpha}{\cos
\beta}-\frac{\cos ^{3}\alpha}{\sin \beta})^{2}]
\nonumber \\
& &\times f_{1}(m_{H}^{2}),
 \nonumber \\
  \end{eqnarray}

\begin{equation}
\Pi _{HH}^{(3)}=\frac{1}{v}(\frac{\cos ^{3}\alpha}
{\sin \beta}-\frac{\sin ^{3}\alpha}{\cos \beta})T_{H}
+\frac{1}{v}\frac{\sin 2\alpha }{\sin 2\beta }
\cos (\alpha -\beta )T_{h}.
\end{equation}
\vskip0.5cm
\begin{center}
{\bf Appendix D}
\end{center}

The mixing diagrams between $H$ and $h$ are also
of the types of Fig. 2. They are again
decomposed into three terms
\begin{equation}
\Pi _{hH}(p^{2})=\Pi _{hH}^{(1)}+\Pi _{hH}^{(2)}
+\Pi _{hH}^{(3)},
\end{equation}
where each term in (52) are given by
\begin{eqnarray}
\Pi _{hH}^{(1)}&=&\frac{1}{2v^{2}}(2m_{h}^{2}
+m_{H}^{2})(2m_{H}^{2}+m_{h}^{2})
(\frac{\sin 2\alpha}{\sin 2\beta})^{2}
\sin (2\alpha -2\beta)
f_{2}(p^{2}, m_{h}^{2}, m_{H}^{2})
   \nonumber \\
& &+\frac{3m_{h}^{2}}{v^{2}}(m_{h}^{2}+\frac{1}{2}
m_{H}^{2})
(\frac{\cos ^{3}\alpha}{\cos \beta}+\frac{\sin ^{3}
\alpha}{\sin \beta})
\frac{\sin 2\alpha }{\sin 2\beta}\sin (\alpha -\beta)
           \nonumber \\
& & \times f_{2}(p^{2}, m_{h}^{2},  m_{h}^{2})
         \nonumber \\
& &+\frac{3m_{H}^{2}}{v^{2}}(m_{H}^{2}+\frac{1}{2}
m_{h}^{2})
(\frac{\cos ^{3}\alpha}{\sin \beta}-\frac{\sin ^{3}
\alpha}{\cos \beta})
\frac{\sin 2\alpha }{\sin 2\beta}\cos (\alpha -\beta)
                  \nonumber \\
& & \times  f_{2}(p^{2}, m_{H}^{2},  m_{H}^{2})
    \nonumber \\
& &-\frac{3m_{h}^{2}m_{H}^{2}}{4v^{2}}\sin (2\alpha
 -2\beta)f_{2}(p^{2}, 0, 0)\nonumber \\
& &+\{\frac{m_{h}^{2}}{v}
(\frac{\sin ^{2}\beta \cos \alpha
}{\cos \beta}+\frac{\sin \alpha \cos ^{2}\beta}{\sin
\beta})+\frac{2m_{G}^{2}}{v}\cos (\alpha -\beta)\}
            \nonumber \\
& &\times \{\frac{m_{H}^{2}}{v}(\frac{\cos ^{2}\beta
\cos \alpha }{\sin \beta}-
\frac{\sin \alpha \sin ^{2}\beta}
{\cos \beta})-\frac{2m_{G}^{2}}{v}\sin (\alpha -\beta)\}
            \nonumber \\
& & \times f_{2}(p^{2},  m_{G}^{2},  m_{G}^{2})
\nonumber \\
& &+\frac{1}{2}\{\frac{m_{h}^{2}}{v}(\frac{\sin ^{2}
\beta \cos \alpha
}{\cos \beta}+\frac{\sin \alpha \cos ^{2}\beta}{\sin
\beta})+\frac{2m_{A}^{2}}{v}\cos (\alpha -\beta)\}
            \nonumber \\
& &\times \{\frac{m_{H}^{2}}{v}(\frac{\cos ^{2}\beta
\cos \alpha }{\sin \beta}-\frac{\sin \alpha \sin ^{2}\beta}
{\cos \beta})-\frac{2m_{A}^{2}}{v}\sin (\alpha -\beta)\}
            \nonumber \\
& & \times f_{2}(p^{2},  m_{A}^{2},  m_{A}^{2})
\nonumber \\
& &+\frac{1}{v^{2}}(m_{h}^{2}-m_{G}^{2})(m_{H}^{2}-
m_{G}^{2})\sin (2\alpha -2\beta )f_{2}
(p^{2},  m_{G}^{2},0)
\nonumber \\
& &+\frac{1}{2v^{2}}(m_{h}^{2}-m_{A}^{2})(m_{H}^{2}-
m_{A}^{2})\sin (2\alpha -2\beta )f_{2}
(p^{2},  m_{A}^{2},0),
\nonumber \\
\end{eqnarray}
\begin{eqnarray}
\Pi _{hH}^{(2)}&=&\{\frac{m_{h}^{2}}{2v^{2}}(\frac{\cos
\alpha \sin \beta}{\cos ^{2}\beta}+\frac{\sin \alpha
\cos \beta}{\sin ^{2} \beta})\sin 2\alpha
\sin (\alpha-\beta)
                       \nonumber \\
& &+\frac{m_{H}^{2}}{2v^{2}}(\frac{\cos
\alpha \cos \beta}{\sin ^{2}\beta}-\frac{\sin \alpha
\sin \beta}{\cos ^{2} \beta})\sin 2\alpha
\cos (\alpha-\beta)
                       \nonumber \\
& &-\frac{m_{G}^{2}}{v^{2}}\sin (2\alpha -2\beta)
\}f_{1}(m_{G}^{2})      \nonumber \\
& &+\{\frac{m_{h}^{2}}{4v^{2}}(\frac{\cos
\alpha \sin \beta}{\cos ^{2}\beta}+\frac{\sin \alpha
\cos \beta}{\sin ^{2} \beta})\sin 2\alpha
\sin (\alpha-\beta)
                       \nonumber \\
& &+\frac{m_{H}^{2}}{4v^{2}}(\frac{\cos
\alpha \cos \beta}{\sin ^{2}\beta}-\frac{\sin \alpha
\sin \beta}{\cos ^{2} \beta})\sin 2\alpha
\cos (\alpha-\beta)
                       \nonumber \\
& &-\frac{m_{A}^{2}}{2v^{2}}\sin (2\alpha -2\beta)
\}f_{1}(m_{A}^{2})      \nonumber \\
& &+\{\frac{3m_{h}^{2}}{4v^{2}}(\frac{\sin 2\alpha}
{\sin 2\beta})^{2}\sin 2(\alpha -\beta)
                    \nonumber \\
& &+\frac{3m_{H}^{2}}{2v^{2}}\frac{\sin 2\alpha}
{\sin 2\beta}(\frac{\cos ^{3}\alpha}{\sin \beta}
-\frac{\sin ^{3}\alpha}{\cos \beta})\cos (\alpha
-\beta)\}f_{1}(m_{H}^{2})
                    \nonumber \\
& &+\{\frac{3m_{H}^{2}}{4v^{2}}(\frac{\sin 2\alpha}
{\sin 2\beta})^{2}\sin 2(\alpha -\beta)
                    \nonumber \\
& &+\frac{3m_{h}^{2}}{2v^{2}}\frac{\sin 2\alpha}
{\sin 2\beta}(\frac{\cos ^{3}\alpha}{\cos \beta}
+\frac{\sin ^{3}\alpha}{\sin \beta})\sin (\alpha
-\beta)\}f_{1}(m_{h}^{2}),
                    \nonumber \\
\end{eqnarray}
\begin{eqnarray}
\Pi _{hH}^{(3)}&=&\frac{1}{v}\frac{\sin 2\alpha }
{\sin 2\beta }\{\cos (\alpha-\beta)T_{H} +\sin (
\alpha -\beta)T_{h}\}.   \nonumber \\
\end{eqnarray}

\pagebreak
\begin{center}
{\bf References}
\end{center}
\begin{description}
\item{[1]}
J.F. Gunion,  H.E. Haber, G. Kane and S. Dawson,
{\it The Higgs Hunter's Guide} (Addison-Wesley
Publishing  Company, 1990).
\item{[2]} M. Veltman, Acta. Phys. Pol. {\bf B 8},
  475 (1977);  Phys. Lett. {\bf 70 B}, 253 (1977).
\item{[3]} M. Einhorn and J. Wudka, Phys. Rev.
{\bf D 39}, 2758  (1989); {\it ibidem} {\bf D 47}, 5029
 (1993).
\item{[4]}
M. Veltman,  Nucl. Phys. {\bf B 123},  89 (1977);
P.  Sikivie,  L.  Susskind, M.  Voloshin and V.
Zakharov,  Nucl.  Phys.  {\bf B 173},  189 (1980);
M.B. Einhorn,D.R.T. Jones and M. Veltman,
Nucl. Phys.  {\bf B 191}, 146 (1981).
\item{[5]} D. Toussaint, Phys. Rev. {\bf D 18},
1626 (1977).
\item{[6]}
S. Bertolini,  Nucl.  Phys. {\bf B 272},  77 (1986);
W. Hollik,  Z. Phys.  {\bf C 32}, 291 (1986);
{\it ibidem} {\bf
C 37},   569 (1988).
\item{[7]}
A. Denner, R.J. Guth, and J.H. K{\" u}hn,  Phys. Lett.
{\bf B 240}, 438 (1990).
\item{[8]}
M. Peskin and T. Takeuchi, Phys. Rev. Lett. {\bf 65}
, 964 (1990);   Phys. Rev. {\bf D 46}, 381 (1992);
G. Altarelli and R. Barbieri, Phys. Lett. {\bf B
253 }, 161 (1991).
\item{[9]}
T. Inami, C.S. Lim and A. Yamada,  Mod. Phys. Lett.
  {\bf A7}, 2789 (1992).
\item{[10]}
C.D. Froggatt, R.G. Moorhouse and I.G. Knowles,
Phys. Rev. {\bf D 45}, (1992) 2471;  Nucl. Phys.
{\bf B 386}, 63 (1992).
\item{[11]}
G.L.  Kane and C.-P. Yuan, Phys. Rev.  {\bf D 49},
  2231 (1989);
V. Barger, K. Cheung, T. Han and D. Zeppenfeld,
 Phys. Rev. {\bf D 48}, 5433, (1993).
\item{[12]}
S. Dawson and S. Willenbrock,  Phys. Rev. Lett.
{\bf 62}, 1232 (1989); Phys. Rev. {\bf D 40}, 2880
(1989);
W.J. Marciano and S.S.D. Willenbrock, Phys. Rev.
{\bf D 37}, 2509 (1988).
\item{[13]}
S.N. Gupta, J.M. Johnson  and  W.W. Repko,
Phys. Rev. {\bf D 48}, 2083 (1993).
\item{[14]}
L. Durand,  J.M. Johnson and J.L. Lopez, Phys. Rev.
{\bf D 45}, 3112 (1992);
P.N. Maher, L. Durand  and K. Riesselmann,
  Phys. Rev. {\bf D 48}, 1061 (1993).
\item{[15]}
V. Barger,  K. Cheung, T. Han, and D. Zeppenfeld,
  Phys. Rev.  {\bf D 44}, 2701 (1991); {\bf D 48},
5433 (1994);
M.D. Hildreth,  T.L. Barklow and D.L. Burke,
Phys. Rev.  {\bf D 49} ,  3441 (1994).
\item{[16]}
S. Kanemura, T. Kubota and E. Takasugi,  Phys.
Lett. {\bf B 313},  155 (1993);
J.  Maalampi, J.  Sirkka and I.  Vilja,
  Phys. Lett. {\bf B 265}, 371 (1991).
\item{[17]}
B.W. Lee, C. Quigg and H.B. Thacker, Phys. Rev.
{\bf D 16}, 1519 (1977).
\item{[18]}
D. Kominis and R.S. Chivukula,  Phys. Lett.
{\bf B 304},  152 (1993).
\item{[19]}
H. Georgi,  Hadr. J.  Phys.  {\bf 1},  155 (1978).
\item{[20]}
S.L. Glashow and S. Weinberg, Phys. Rev. {\bf D 15},
1958 (1977);
J. Liu and L. Wolfenstein,  Nucl. Phys. {\bf B 289},
1 (1987).
\item{[21]}
L. McLerran, M. Shaposhnikovm N. Turok and M. Voloshin,
Nucl. Phys. {\bf B 256},  451 (1991);
A.G. Cohen, D.B. Kaplan and A.E. Nelson,  Phys. Lett.
 {\bf B 263},  86 (1991);
A.E. Nelson, D.B. Kaplan and A,G. Cohen,  Nucl. Phys.
 {\bf B373}, 453 (1992).
\item{[22]}
G.  Cvetic,   Phys. Rev.  {\bf D 48},  5280 (1993).
\item{[23]}
R. Casalbuoni, D. Dominici, F. Feruglio and R. Gatto,
Nucl. Phys. {\bf B 299}, 117 (1988);   Phys. Lett. {\bf
B 200}, 495 (1988);   R. Casalbuoni, D. Dominici, R.
Gatto and C. Giunti, Phys. Lett. {\bf B 178}, 235
(1986).
\item{[24]}
J.M. Cornwall, D.N. Levin and  G. Tiktopoulos,
  Phys. Rev.  Lett. {\bf 30}, 1268 (1973); Phys.
  Rev.  {\bf D 10},  1145 (1974).
\item{[25]}
J.C. Taylor, {\it Gauge Theories of Weak Interactions}
  (Cambridge University Press, Cambridge, England,
(1976)) Chap. 14.
\item{[26]}
A.  Czarnecki,  Phys. Rev.  {\bf D 48},  5250
(1993).
\item{[27]}
J. Fujimoto,  M. Igarashi,  N. Nakazawa,
Y. Shimizu and K.  Tobimatsu,
  Prog. Theor. Phys. Suppl.
{\bf 100}, 1 (1990).
\item{[28]}
V. Barger,  J.L.  Hewett and R.J.N. Phillips,
Phys. Rev.  {\bf D 41}, (1990) 3421;
A. J. Buras, P.  Krawczyk,  M.E. Lautenbacher,
and C.  Salazar,  Nucl. Phys.  {\bf B 337},
 284 (1990);
J.F. Gunion and B. Grzadkowski,  Phys.
Lett. {\bf B 243},  301 (1990);
D. Cocolicchio and J.-R. Cudell,  Phys. Lett.
 {\bf B245},  591 (1990);
J. Rosiek,  Phys. Lett. {\bf B 252},  135 (1990).
\item{[29]}
S.  Kanemura and T. Kubota, in preparation.
\item{[30]}
Y. Okada, M. Yamaguchi,  and T. Yanagida,
Prog. Theor. Phys.  {\bf 85},  1 (1991);
J. Ellis,  G.  Ridolfi and F.  Zwirner, Phys. Lett.
{\bf B 257},  83 (1991);
H.E. Haber  and   R. Hempfling,  Phys. Rev.  Lett.
{\bf 66},  1815 (1991);
P.H. Chankowski, J. Rosiek  and S. Pokorski,  Phys.
 Lett.  {\bf B 274}, 191 (1992).

\end{description}

\vfill\eject
\begin{flushleft}
{\bf Table 1}
\end{flushleft}
\noindent
Combinations of internal particles $(X, Y)$
running
in Fig. 2(a) and $X$ in Fig. 2(b) for
$\Pi _{ww}^{(i)}$ and $\Pi _{wG}^{(i)}$
  $(i=1,2)$.
\begin{center}
\begin{tabular}{c|l}       \hline
Propagator  &  Internal particle species  \\  \hline
$\Pi _{ww}^{(1)}$       &
     $(h,w)$,  $(H,w)$,  $(h,G)$,  $(H,G)$,  $(A,G)$ \\
$\Pi _{ww}^{(2)}$       &
     $G$, $A$, $h$, $H$                   \\
$\Pi _{wG}^{(1)}$       &
     $(h,w)$,  $(H,w)$,  $(h,G)$,  $(H,G)$           \\
$\Pi _{wG}^{(2)}$       &
     $G$,  $A$, $h$, $H$                   \\
          \hline
\end{tabular}
\end{center}
\vskip3cm

\begin{flushleft}
{\bf Table 2}
\end{flushleft}
\noindent
Combinations of internal particles $(X, Y)$
running  in Fig. 2(a) and $X$ in Fig. 2(b) for
$\Pi _{HH}^{(i)}$ and $\Pi _{hH}^{(i)}$
  $(i=1,2)$.

\begin{center}
\begin{tabular}{c|l}       \hline
Propagator  &  Internal particle species  \\  \hline
$\Pi _{HH}^{(1)}$,  $\Pi _{hH}^{(1)}$      &
     $(H,H)$,  $(h,H)$,  $(h,h)$,  $(w,w)$,  $(z,z)$, \\
  &  $(G,G)$,  $(A,A)$,  $(A,z)$,  $(G,w)$           \\
$\Pi _{HH}^{(2)}$,  $\Pi _{hH}^{(2)}$       &
     $G$, $A$, $h$, $H$                   \\
          \hline
\end{tabular}
\end{center}
\vskip1cm

\vfill\eject
\begin{flushleft}
{\bf Table 3}
\end{flushleft}
\noindent
Combinations of internal particles $(X, Y,; Z)$
in Fig. 4(a) and $(X, Y)$ in Figs. 4(b)
and 4(c)for the vertices  $\Gamma ^{(i)}$
  $(i=1,2,3)$.

\begin{center}
\begin{tabular}{c|l}           \hline
Vertex  &   Internal particle species   \\   \hline
$\Gamma ^{(1)}$   &
       $(w,w;h)$,  $(w,w;H)$,  $(G,G;h)$,  $(G,G;H)$,  \\
  &    $(G,G;A)$,  $(A,A;G)$,  $(G,w;h)$,  $(w,G;h)$,  \\
  &    $(G,w;H)$,  $(w,G;H)$,  $(H,H;w)$,  $(H,H;G)$,  \\
  &    $(H,h;w)$,  $(h,H;w)$   $(H,h;G)$,  $(h,H;G)$,  \\
  &    $(h,h;w)$,  $(h,h;G)$                          \\
$\Gamma ^{(2)}$   &
       $(w,w)$,  $(z,z)$,  $(G,G)$,  $(A,A)$,  $(A,z)$,\\
  &    $(G,w)$,  $(H,H)$,  $(H,h)$,  $(h,h)$          \\
$\Gamma ^{(3)}$    &
       $(w,h)$,  $(w,H)$,  $(G,h)$,  $(G,H)$,  $(G,A)$ \\
                         \hline
 \end{tabular}
\end{center}

\pagebreak
\begin{center}
{\bf Figure Captions}
\end{center}
\begin{description}
\item{Fig. 1}

Tadpole diagrams of $H$ and $h$ fields
contributing to (a) $T_{H}$ and (b) $T_{h}$.
\item{Fig. 2}

Self-Energy diagrams contributing to
(a) $\Pi _{ij}^{(1)}$, (b) $\Pi _{ij}^{(2)}$, and
(c) $\Pi _{ij}^{(3)}$, respectively.
The diagram (c) denotes the counter terms,
which are expressed by the tadpole
contributions.
A pair of indices  $(i,j)$  refers to
either of $(w,w)$,
 $(w,G)$,  $(H,H)$, or $(h,H)$.
Internal particle species are given in Tables 1 and 2.
\item{Fig. 3}

General configuration of the $Hww^{\dag }$
vertex. The Nambu-Goldstone bosons are put on the
 mass shell ($p_{1}^{2}=p_{2}^{2}=0$), while the
the Higgs boson $H$ are kept off-shell to keep
generality.

\item{Fig. 4}

Radiative corrections to the $Hw^{\dag}w$ vertex
contributing to (a) $\Gamma ^{(1)}$,
 (b) $\Gamma ^{(2)}$  and
(c) $\Gamma ^{(3)}$.
  Internal particle species $(X,Y; Z)$
in (a) and $(X,Y)$  in (b) and those in
(c) are given in Table 3.

\item{Fig.    5}

The decay width (35) as a function of $m_{G}$.
The mixing angles are determined  by $\tan \alpha =
\tan \beta =2$. The masses of the neutral
Higgs bosons
 are assumed to be $m_{H}=300$ GeV,
 $m_{h}=400 $ GeV. The CP-odd Higgs boson
mass is taken as  (a) $m_{A}=350 $ GeV
(solid line),   (b) $m_{A}=700$  GeV
(dashed line)  and (c) $m_{A}=1000 $
 GeV (dotted line), respectively.
\pagebreak
\item{Fig.   6}

The decay width (35) as a function of $m_{G}$.
The mixing angles are determined  by $\tan \alpha =
\tan \beta =10$. The masses  of the neutral
Higgs bosons
are assumed to be $m_{H}=300 $ GeV,
$m_{h}=400 $ GeV.  The CP-odd Higgs boson mass
is taken as   (a) $m_{A}=350 $ GeV (solid
line),    (b) $m_{A}=700 $ GeV (dashed
line) and (c) $m_{A}=1000 $ GeV
(dotted line), respectively.

\item{Fig.  7}

The decay width (35) as a function of $m_{G}$.
The mixing angles are determined  by $\tan \beta =
2 $ and $\sin ^{2}(\alpha -\beta)=1$.
 The masses  of the  neutral Higgs bosons
are assumed to be $m_{H}=300 $ GeV,
$m_{h}=400$  GeV.  The CP-odd Higgs boson mass
 is taken as (a) $m_{A}=350$  GeV (solid line),
  (b) $m_{A}=700 $ GeV (dashed line) and
 (c) $m_{A}=1000 $ GeV  (dotted line),
 respectively.

\item{Fig.  8}

The decay width (35) as a function of $m_{G}$.
The mixing angles are determined  by $\tan \beta =
10 $ and $\sin ^{2}(\alpha -\beta)=1$.
 The masses  of the  neutral Higgs bosons are
assumed to be
$m_{H}=300 $ GeV,  $m_{h}=400$  GeV.  The CP-
odd Higgs boson mass is taken as (a) $m_{A}=350$
 GeV (solid line), (b) $m_{A}=700$  GeV
(dashed line) and (c) $m_{A}=1000  $ GeV
(dotted line), respectively.

\end{description}
\end{document}